\documentclass{emulateapj}
\usepackage{amsmath}
\usepackage{epsfig}
\usepackage{apjfonts}
\usepackage{epsfig}

\newcommand{\beq}{\begin{equation}}
\newcommand{\eeq}{\end{equation}}
\def\gtsim {\lower .1ex\hbox{\rlap{\raise .6ex\hbox{\hskip .3ex
        {\ifmmode{\scriptscriptstyle >}\else
                {$\scriptscriptstyle >$}\fi}}}
        \kern -.4ex{\ifmmode{\scriptscriptstyle \sim}\else
                {$\scriptscriptstyle\sim$}\fi}}}
\def\Vmax{V_{\rm max}}

\def\rcore{r_{\rm core}}
\def\rg{r_{\gamma}}
\def\Msun{{\rm M_\odot}}

\def\kms{{\rm km}\,{\rm s}^{-1}}
\def\LCDM{$\Lambda$CDM}

\shorttitle{Large core in Fornax?}
\shortauthors{Strigari et~al.}

\begin{document}

\title{A large dark matter core in the Fornax dwarf spheroidal galaxy?}

\author{Louis E. Strigari\altaffilmark{1,2}
 James S. Bullock\altaffilmark{1},
 Manoj Kaplinghat\altaffilmark{1}, 
Andrey V. Kravtsov  \altaffilmark{3},
   Oleg Y. Gnedin\altaffilmark{4},
 \\ Kevork Abazajian\altaffilmark{5}, 
 Anatoly A. Klypin\altaffilmark{6} 
 \\
}
\altaffiltext{1}{Center for Cosmology,
Dept. of Physics \& Astronomy, University of California, Irvine, CA; lstrigar@uci.edu}
\altaffiltext{2}{McCue Fellow}
\altaffiltext{3}{Dept. of Astronomy \& Astrophysics, KICP, Enrico Fermi Institute, 5640 S. Ellis Ave., The University of Chicago, Chicago, IL 60637} 
\altaffiltext{4}{Astronomy Department, The Ohio State University, Columbus, OH}
\altaffiltext{5}{Theoretical Division, MS B285, Los Alamos National Laboratory, Los Alamos, NM 87545}
\altaffiltext{6}{Astronomy Dept, New Mexico State University, Las Cruces, NM 88001}

\begin{abstract}

We use measurements of the stellar  velocity dispersion profile of the
Fornax  dwarf spheroidal galaxy   to  derive constraints on its   dark
matter distribution. Though the data are unable to distinguish between
models with small cores  and those with cusps,  we  show that a  large
$\gtrsim 1$ kpc dark matter core in Fornax is highly implausible.  Irrespective of the
origin of  the core, reasonable  dynamical  limits on  the mass of  the
Fornax halo constrain its core radius  to  be no larger  than  $\sim
700$~pc.  We   derive an upper limit  of   $\rcore \lesssim  300$~pc by
demanding that  the central phase-space density  of Fornax  not exceed
that  directly inferred from   the rotation curves of low-mass  spiral
galaxies. Further, if the halo is composed of warm dark matter then phase-space
constraints force  the  core  to be  quite  small  in order  to  avoid
conservative  limits  from  the  Ly$\alpha$  forest power  spectrum,
$\rcore  \lesssim~85$~pc. We  discuss  our results  in the context  of the
idea that   the  extended globular    cluster
distribution  in  Fornax can be explained  by  the presence of a large
$\sim  1.5$~kpc core. A  self-consistent core  of this  size would be
drastically inconsistent with  the expectations of  standard warm or cold dark
matter models, and  would also require  an unreasonably massive
dark matter halo, with $\Vmax \simeq 200~\kms$.

\end{abstract}

\keywords{
cosmology: dark matter, cosmology: observations, cosmology: theory, 
galaxies: kinematics and dynamics, galaxies: structure, galaxies: formation, dwarfs: galaxies}

\section{Introduction}
\label{sec:introduction}

The   central density   distribution in  many  dark matter   dominated
galaxies  appears   to be  lower  than what   is expected  in standard
$\Lambda$ + Cold Dark Matter (\LCDM) models
\citep{moore94,flores_primack94,deBlok01,zentner_02,kuhlen_etal05,simon_etal05,zackrisson06}.
This discrepancy, along with other potential difficulties
\citep[e.g.][]{klypin99,moore99}, 
can be ameliorated by considering alternative dark matter models
\citep[e.g.][and references therein]{zentner_03}.  One
particularly  intriguing scenario is   Warm  Dark Matter (WDM),  which
differs from CDM in that the dark matter has a non-negligible velocity
dispersion
\citep{bond_etal,blumenthal_etal82,pagels_primack82,dodelson_widrow94,hogan_dalcanton01,abazajian06}. 
Generally,   these WDM models   come with  two  distinct observational
signatures. First, there  is a reduction in  clustering and a delay in
collapse  times for structures  on linear scales approximately smaller
than  the free-streaming  scale of  the WDM particle.   This effect is
currently  probed  by   measurements of  the Ly$\alpha$   forest power
spectrum and this places strong   constraints  on the mass of   the WDM
particle   \citep{viel_etal05,abazajian06a,seljak06}.   Second,  the velocity dispersion of  WDM  imposes an upper  limit on
the phase-space density,  defined as $Q \equiv \rho/\sigma^{3}$, where
$\rho$ is the  density and $\sigma$ is the  velocity dispersion of the
dark matter. The upper  limit on Q implies  that the density profiles
of WDM halos must saturate to form a constant density core at an inner
radius defined by  $Q < Q_{\rm max}$  \citep{tg79}.  This  is in stark
contrast to CDM predictions that give dark halo density profiles
that  rise  steeply towards  the center, $\rho
\propto r^{\gamma}$, with $\gamma \sim 1$,
\citep[e.g.][]{dc91,nfw96,moore99,klypin_etal01,navarro_etal04,diemand_etal05} 
and have divergent phase-space 
profiles $Q  \propto  r^{-1.9}$ \citep{tn01}. 
 
Dwarf  spheroidal  galaxies   (dSph's)  are  diffuse,   low-luminosity
systems, with a total  mass believed to be  dominated by their host dark
matter halos  \citep{mateo_98}.  Assuming  that  these galaxies are  in
dynamical equilibrium, the stars  act as tracers of  the gravitational
potential,  and can be   used as a  probe  of the dark matter  density
profile.  Indeed, dSph's provide   a  unique testing ground  for   the
nature  of dark matter, because in these low-mass  systems the phase-space cores are the most observationally accessible \citep*{sellwood00,dalcanton_hogan01,abw02}.  Unfortunately,
attempts  to constrain uniquely the dark  matter profiles of dwarfs using stellar  velocity
dispersion data are plagued  by degeneracies. 
We emphasize here that while the data are  currently
unable to settle the question of a central density core versus cusp,
they do provide powerful constraints on the combination of
core size ($\rcore$) and halo maximum circular velocity ($\Vmax$).
Each solution maps directly to a constraint on $Q$.

\begin{figure*}
\epsscale{0.8}
\plotone{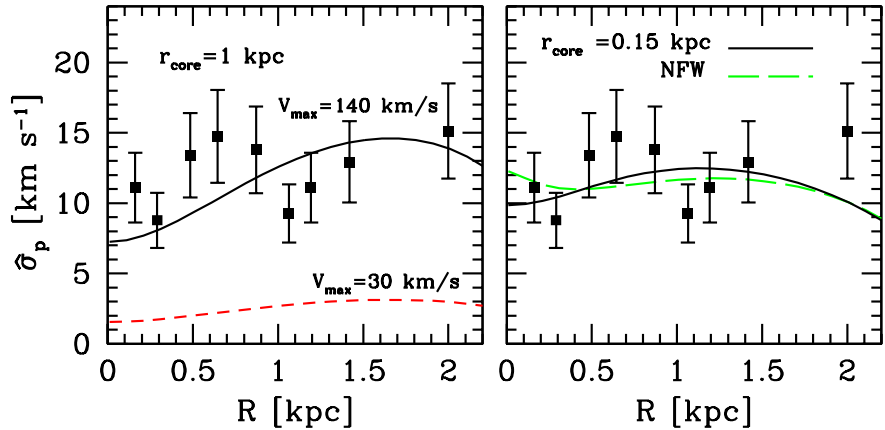}
\caption{Measurements of the Fornax velocity dispersion profile compared to models for the dark matter halo. {\it Left panel}:  Predictions for the case of a fixed $\rcore = 1$ kpc; the solid curve shows a large $\Vmax$ = 140 km/s, and the dashed curve shows a small $\Vmax$ = 30 km/s.  
{\it Right panel}: The best-fitting cases for two distinct density profiles: the solid curve shows the fit for the $\alpha =1.5$ profile described in the text with $\Vmax = 28$ km/s, and the long-dashed curve shows the best-fitting cuspy NFW profile with $\Vmax = 50$ km/s. 
\label{fig:sigma}}
\end{figure*}

The Fornax dSph is approximately $138$~kpc from the Milky Way \citep[e.g.,][]{buonanno_etal99}, 
in the vicinity of its orbital pericenter
\citep{dinescu_etal04}.  Its stellar population is dominated by
intermediate  age stars ($\sim 4$~Gyr), 
although Fornax has also both
very young stars ($\sim 0.2-2$~Gyr) and very old stars \citep[$\gtrsim
10$~Gyr, e.g.,][and references  therein]{pont_etal04}.
Among the most puzzling observed properties of Fornax is the fact
that it hosts 5  globular clusters distributed over a wide range
of distances from the center of this galaxy.  Standard arguments suggest that
(in the absence of external heating) these systems should have sunk to
the center via dynamical friction in much less than a Hubble time
\citep{oh_etal00}. Recently,
\citet[][G06              hereafter]{goerdt_etal06}                and
\citet{sanchezsalcedo_etal06} suggested that the presence of these
clusters provides good evidence for the existence of a large, constant
density dark  matter core in the  center of Fornax.  Specifically, G06
argued that the globular clusters  will stall in their orbital  decay
if they  encounter  a constant density  core.  Thus the   globular clusters should   be located in  a narrow
range of radii  from the center  of  Fornax corresponding to its  halo
core size.  The  Fornax globular clusters  are observed to  orbit at a
range of  {\it   projected} distances from  the  center  of galaxy  of
between 0.24 and 1.60 kpc
\citep{mackey_gilmore03}.  This suggests that a core size
$\rcore \gtrsim 1.5$~kpc would be required to explain the observed 
distribution.  
Another idea, which relies on a somewhat
fine-tuned timing argument,  is that only the central-most globular cluster is stalled at the core, $\rcore \gtrsim 240$ pc, (G06; J. Read, private communication) and the other four globular clusters were
simply formed at a large enough radius that they have not sunk to the center.

In this paper we consider the general plausibility  of a large core in
Fornax.  In section \ref{sec:dynamical},  we focus on the allowed dark
halo solutions which reproduce the observed stellar   kinematics, and we present a toy model
to understand the nature of the these solutions. In
section  \ref{sec:results}, we combine these  solutions with limits on
the  WDM  mass  from the  Ly$\alpha$  forest  power spectrum  to place
constraints on the  maximal core size. We   also  discuss
additional, more independent constraints on the core size. We conclude
by discussing these results in  the  context of the observed  globular
cluster distribution in Fornax.

\begin{figure*}
\epsscale{0.8}
\plotone{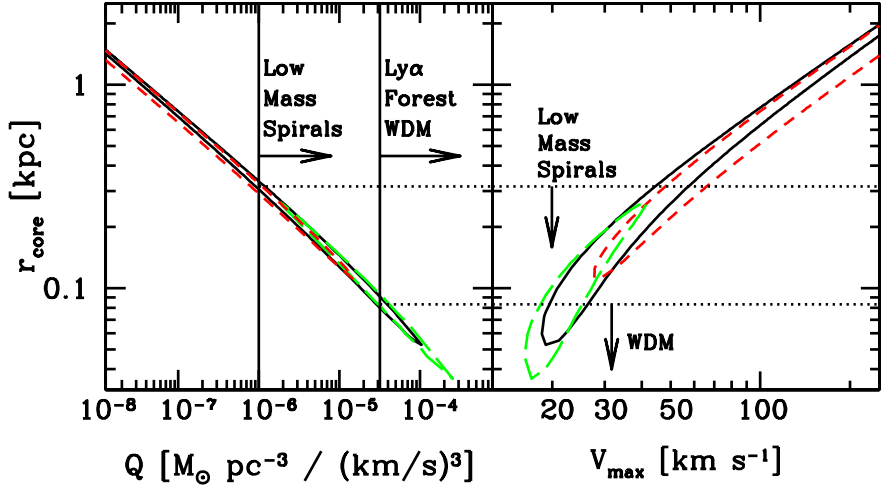}
\caption{\label{fig:rcore} 
Constraints on  the core radius   of Fornax as   a function of central
phase-space  density (left)  and   maximum circular velocity   (right)
derived from  the velocity dispersion  profile.  We define $\rcore$ as
the  radius where the log-slope  of the density   profile is $\gamma =
-0.1$ in  equation \ref{eqt:rho} with $\alpha  = 1.5$,  $\rcore \simeq
0.1 r_0$.   The long-dashed,  solid, and short-dashed lines  use 
$\beta=0.5$, $0.0$,  and $-0.5$ respectively.  Values
of  $\rcore$ with $Q \gtrsim  3 \times 10^{-5}$ M$_{\odot} $ pc$^{-3}$
$(\rm{km/s})^{-3}$ are ruled  out by Ly$\alpha$ forest  constraints in
the case of WDM with g = 2 ($\rcore \lesssim  85$~pc).  In a  more  general class of dark matter
models, directly observed phase-space limits from 
low-mass spiral rotation curves, $Q  \gtrsim   10^{-6}$
M$_{\odot}  $ pc$^{-3}$  $(\rm{km/s})^{-3}$,
demand $\rcore \lesssim 300$~pc. 
Most generally and irrespective of the cause of the core,
a Fornax halo with $\Vmax \gtrsim
100~\kms$ is disfavored from dynamical considerations.  This implies
$\rcore \lesssim 700$~pc.  }
\end{figure*}

\section{Dynamical constraints on the dark halo of Fornax}
\label{sec:dynamical}

For a stellar  system in equilibrium  and
embedded  in  a   spherically  symmetric  dark    matter   potential, 
the radial  stellar velocity dispersion $\sigma_r(r)$  is
given by the Jeans equation:
\beq
\label{eq:jeans}
r \frac{d(\rho_{\star} \sigma_r^2)}{dr} =  - \rho_{\star}(r) V_c^2(r)
        - 2 \beta(r) \rho_{\star} \sigma_r^2.
\eeq
Here the circular velocity is governed by the mass distribution
 $V_{\rm c}^2(r) = G M(r)/r$ and the anisotropy parameter is 
$\beta \equiv 1 - \sigma_{\theta}^2/\sigma_r^2$, where $\sigma_r$ and $\sigma_{\theta}$ are
the radial and angular velocity dispersions, respectively. We consider $\beta (r)$ to be constant in radius, and explore three  choices  of $\beta  =-0.5$, $0$, and $0.5$, using $\beta = 0$  as our fiducial case.  The
appropriate value for $\beta$ is unknown, however this range
brackets reasonable  choices.    Dissipationless CDM simulations  show
that   $\beta$ for    dark    matter  increases from    $\sim   0$ to $0.6$
 as the radius increases towards the virial radius of
the halo \citep{lc96,colin00,diemand04,faltenbacher,wojtak_etal}.
However, models of dSph galaxies \cite[e.g.][]{mayer04} suggest that a more
natural value for the stars may be negative, $\beta \sim -0.5$.
Our main results are not sensitive to the choice of $\beta > 0$, 
and we note that $\beta < 0$ provides an even stronger upper limit on the core size than 
the $\beta=0$ case.

We assume that the stellar component and the dark matter component are
uncoupled, each with an independent density distribution, and that 
$M(r)$ is dominated by the dark matter. We model the
stellar distribution,   $\rho_*(r)$, as a   spherically symmetric King
profile (King 1962) and adopt $r_{\rm c}  = 0.39$~kpc and $r_{\rm t} =
2.7$~kpc for its King core and tidal  radii \citep{mateo_98}.  For the
dark matter, we consider a density profile of the form
\beq
\rho(r)= \frac{\rho_0}{\left[1 + (r/r_0)^\alpha\right]^{3/\alpha}},
\label{eqt:rho}
\eeq
where $\rho_0$ is a central core density and $r_0$ is a characteristic radius.
Note that we demand that the profile falls off at large radius as
$r^{-3}$, as seen in isolated galaxies \citep{prada_etal03}.
The parameter $\alpha$ controls the sharpness of the transition from
flat to falling density with radius.
In our fiducial case we adopt $\alpha = 1.5$.
This choice yields a profile that matches very closely that advocated by
\cite{burkert95} to match rotation curve data for low-mass galaxies.  
However, our conclusions do not change considerably
if we adopt other reasonable choices $\alpha = 1.0$ or $2.0$ (see below).
Note that once $\alpha$ is fixed,
 any two independent parameters in the
mass distribution (e.g. $r_0$ and the maximum of the circular velocity,
$\Vmax$) define the density
profile completely.  The circular velocity curve
implied by equation \ref{eqt:rho}
peaks at a radius $r_{\rm max} = (4.4$, $3.2$, $2.9$)~$r_0$ 
for $\alpha =$ ($ 1.0$, $1.5$, $2.0$).

The log-slope of this profile, $\gamma$, 
gradually approaches zero towards the center, but a constant density
``core radius'' is never achieved at finite $r$.
A natural way to define a core radius in this case 
is to use the radius where $\gamma$ reaches some small
value, such that $\rcore \equiv \rg$, where $\rg$ is defined as follows:
\beq
\rg  =  r_0 \left[ -3/\gamma - 1\right]^{-1/\alpha}.  
\label{eq;logslope}
\eeq
More general forms for $\rho(r)$ may be considered, but
this log-slope definition of core radius is always possible. 
In this paper we define the core radius to be where
$\gamma = -0.1$.  This is the same definition used by G06.
For $\alpha = 1.5$ we have $\rcore \simeq 0.1 r_0$.

In order to compare to measurements of the radial stellar velocity dispersion profile, we
project  equation   \ref{eq:jeans} along  the  line-of-sight to  the
observer.  Performing this projection gives 
\beq
\hat{\sigma}_{p}^{2}(R) = \frac{2}{I(R)} \int_{R}^{\infty} \left ( 1 - \beta \frac{R^{2}}{r^2} \right )
\frac{\rho_{\star} \sigma_{r}^{2} r}{\sqrt{r^2-R^2}} dr, 
\label{eq:proj}
\eeq
where $I(R)$ is the surface density as determined from $\rho_{\star}$, and $R$ is the radial distance
from the center of the galaxy. 
The line-of-sight stellar velocity dispersion was recently measured by \citet{walker_etal05}.   
The authors   utilize  a  primary  sample  of
$N=209$ stars and perform cuts to remove interloping stars, and present  data for   
three  removal schemes  that  yield  $N=179$,
$N=185$, and $N=189$ stars   respectively.  For our analysis,   we use
their $N=185$ sample,  as shown by the  points  with with error bars  in
Figure \ref{fig:sigma}.  Our  conclusions are not strongly affected by
this particular choice.

The lines in Figure \ref{fig:sigma}  show the predicted  $\hat{\sigma}_p$
profiles for four   different input dark matter  density distributions
compared to the data.  In  each case we have  assumed $\beta = 0$.
Each  curve in  the left  panel assumes  $\alpha   = 1.5$ in  equation
\ref{eqt:rho} and fixes the core size to  $\rcore = 1$~kpc.  The solid
line shows the  best-fit case to the data  with these restrictions and
has $\Vmax =   140~\kms$.  The  dashed   line, on  the  other  hand,
corresponds to what  might be considered a  more reasonable halo mass,
with $\Vmax  = 30\ \kms$.  Clearly, in  order to reproduce a reasonable
stellar velocity dispersion profile with  such a large core, the  dark
matter halo  of Fornax must  be quite massive.   In order to match the
data with $\Vmax  = 30~\kms$, a smaller  core must be used.  The solid
line in the right  panel uses the same  profile shape with $\Vmax = 30~\kms$,  
but now with $\rcore =  0.15$~kpc.  There is a clear degeneracy
between  $\Vmax$ and  $\rcore$ --  a  large flat  density distribution
requires a deeper potential well in order to reproduce  the data.  The
dashed  line in  the right hand   panel shows  a best-fit   case for a
(cuspy) NFW dark matter profile.  Note that these data do not exclude 
an inner cusp.

Figure \ref{fig:rcore} illustrates a more general
exploration of the allowed parameter space 
for  $\alpha = 1.5$.   We
demand that  the  allowed  fit region obey
 $\chi^{2} <
\chi^{2}_{{\rm min}}+\Delta  \chi^{2}$,  where $\Delta  \chi^2 =  4.6$
defines the $90  \%$  confidence level  for a  $\chi^{2}$ distribution
with   two  degrees of freedom.   The parameters  $\rcore$ and $\Vmax$
define the density   profile and   must  lie  within the  
banana-shaped contour in the right panel in order to
reproduce the  observed $\hat{\sigma}_{p}$.   At large $\Vmax$   the
contours follow $\rcore \propto \Vmax$ and we discuss  how this can be
understood analytically  below.     The left  panel shows
the same best-fit parameter space,  now expressed in terms of $\rcore$
and  the  central ($r=0$) phase-space  density of the dark matter, $Q$,  
implied by each solution. The  different line  types in   Figure \ref{fig:rcore} correspond   to
different assumed values for  the velocity anisotropy: $\beta  = -0.5$
(short-dash), $\beta =0$ (solid),  and $\beta = 0.5$ (long-dash). Note
that  for $\beta  =  0.5$ solutions  with $\rcore  \gtrsim 250$~pc and
$\Vmax  \gtrsim 30~\kms$ are  disfavored.  For $\beta =-0.5$ solutions
with $\Vmax \lesssim 30~\kms$ are disfavored.  

Similar results for the velocity anisotropy for Fornax
were uncovered in previous studies \citep{lokas02,kazantzidis_etal04}. In particular \citet{lokas02} 
showed that the degeneracy 
with $\beta$ and the density profile can be broken by considering higher order 
moments of the velocity distribution function. Though not shown, we have
also explored the effects of varying the specific dark halo profile
{\em shape} by varying $\alpha$ in equation \ref{eqt:rho}.
As $\alpha$ increases, the allowed core size increases at fixed $\Vmax$.
For example, at $\Vmax = 30~\kms$ the best-fit value of $\rcore$
varies from $50$~pc to $250$~pc as $\alpha$ varies from $1.0$ to $2.0$.
We discuss the origin of this variation in the next section.

Note that while some large values of $\rcore$ ($\sim$~kpc) are allowed
by the velocity dispersion data, they  require {\it very} large values
of  $\Vmax$ ($\gtrsim 100~\kms$).  One  may be  concerned that the quoted
$\Vmax$ values represent  an unphysical extrapolation to large $r_{\rm
max}$  radii based   on   constraints  derived  by   stellar  velocity
dispersions at  $\sim 2$~kpc.  However,  direct observational evidence
suggests that  this is  a  reasonable extrapolation.   Not only is our
adopted  profile   shape      motivated  by    rotation  curve    fits
\citep{burkert95} but large dark matter halos around isolated galaxies
are    known  to  decline no   faster   than  $r^{-3}$   at  large $r$
\citep{prada_etal03}.  In this case, the only effect that could act to
truncate  this fall-off is tidal striping.   If Fornax is  at a distance 
$D$ from the Milky Way, then the tidal radius may be estimated via the
Jacobi approximation $R_t \simeq D \left(M_{\rm Fornax}(<R_t)/(3
M_{\rm   MW}(<D))\right)^{1/3}$.  For a flat, Milky Way
rotation    curve with $\Vmax = 220~\kms$  we      obtain     $R_t \simeq   80~{\rm    kpc}
\left(\Vmax/(200~\kms)\right)^{2/3}$.     Compare  this to   our
central $\Vmax \simeq  200~\kms$ solution,  which gives $r_{\rm max} =
3.2 \r0 = 32  \rcore \simeq 48$~kpc and  we see that the peak rotation
curve should be little affected by tidal truncation.  We conclude that
while a large Fornax halo may  be implausible on physical grounds (see
below) the derived constraint itself is meaningful.

It is also possible that the large $\Vmax$ end of the contour
is affected by the  presence  of
interloping  stars, and that the contours actually close at a 
smaller value of $\Vmax$ (and $\rcore$). 
In particular,    removal of interlopers   is most
important in the outer regions of the galaxy. To examine the effect of
interlopers, we have  performed  an independent fit of    the  data
presented  in \citet{walker_etal05}. We use the  same number of radial
bins,  and in each  bin  we assume that the   distribution of stars in
velocity  follows a  Gaussian plus   a  constant,  where the  constant
accounts for interlopers. The main difference we find is that there is
a  lower  dispersion  in  the  last bin  at   $\sim  2$  kpc than   in
\citet{walker_etal05}, with a larger error  in this bin. These results
do not change the shape of the constraints at low $\Vmax$, but 
disfavor large values: $\Vmax \gtrsim 65~\kms$ and $\rcore \gtrsim 400$~pc
are excluded.  

While uncertainties associated with interlopers stars
 may affect the closure of the contour in  Figure
\ref{fig:rcore} at   large values of  $\Vmax$, the  lower  part of the
contour is  more robust  to these  uncertainties.   Indeed, it is this
region  of the  diagram,  and  the $\Vmax$-$\rcore$  degeneracy itself
which  allows  us to use the   implied  phase-space densities in these
galaxies  to  limit core  sizes.  Before   going  on to discuss  these
limits,   we discuss  the  origin  of this  degeneracy using  analytic
arguments and extend these arguments to  explore the dependency of our
constraints on the assumed underlying dark profile shape. Readers interested
in our resulting constraints on the core size of Fornax may skip ahead to section 3.  

\subsection{Toy Model to Understand the Solutions from Stellar Kinematics} 

Here we show that the scaling of the  contours in the $\Vmax$-$\rcore$
can be understood in  the  context of  a toy model.   For
simplicity we will consider $\beta=0$.  From equation \ref{eq:jeans} we see that if
the stellar  mass is sub-dominant,  the stellar velocity dispersion is
governed by  the circular velocity curve of  the  dark matter $V_c(r)$
within the stellar radius, $r_\star \simeq  r_c \simeq 0.5$ kpc.  Here
$r_\star$  corresponds to a weighted average  of  the stellar material
over an integral  $\int {\rm d}r  \, \rho_\star(r) V_c^2(r) /r$.  Thus
$r_\star$ will typically be {\em smaller} than the half-mass radius of
the stars.

Consider now a toy-model dark matter halo with a rotation curve given by
\begin{eqnarray}
V_c(r) = \left\{\begin{array}{cll}
	&  a \Vmax \left(\frac{r}{\rcore}\right)  & r < \rcore \\
        &  a \Vmax \left(\frac{r}{\rcore}\right)^{1/2} & \rcore < r < r_{\rm flat}  \\
	&  \, \Vmax  & r > r_{\rm flat} = \rcore/a^2.	   \end{array} \right.
\end{eqnarray}
This corresponds to a density profile with
a constant density at $r < \rcore$.  The profile
transitions to an $\rho \propto r^{-1}$ regime
beyond the core and finally reaches an 
 isothermal distribution, $\rho \propto r^{-2}$, at $r > r_{\rm
   flat}$. We assume that the transitions occur in a small region
 around $\rcore$ and $r_{\rm flat}$ and that the density profile is
 continuous. 
The variable $a \le 1$ parameterizes 
the sharpness of the transition from a core to flat rotation, 
and can be viewed as analogous to the parameter $\alpha$ in equation
\ref{eqt:rho}. 
For $a=1$, the transition is sudden, $r_{\rm core} = r_{\rm flat}$, 
while for $a \ll 1$ the rotation curve rises very gradually and reaches
$\Vmax$ at $r = r_{\rm flat} = \rcore/a^2$.  
Note that the value of the rotation velocity at $r= \rcore$ is
$V_{\rm core} = a \Vmax$, and thus decreases at fixed $\Vmax$ as the transition
is made more gradual.  
Also note that the density in the core of the halo $\rho_0$ is proportional
to $a\Vmax/\rcore$. 

Now consider a stellar distribution with radius $r_\star$ and with
constant density, $\rho_*(r) =$constant,    
embedded  within the gravitationally-dominant dark halo.
If the halo has a very sharp transition from the core region to flat
rotation, $a=1$ and $\rcore = r_{\rm flat}$, then
equation \ref{eq:jeans} implies that the
{\em central} stellar radial  velocity dispersion, 
$\sigma_\star^0 = \sigma_\star(r=0)$, is
\begin{eqnarray}
\sigma_\star^0 =
 \left\{\begin{array}{cll}
  & \Vmax \frac{1}{\sqrt{2}}  \left(\frac{r_\star}{\rcore}\right) & r_\star \le \rcore \\
 & \Vmax \left[\frac{1}{2} + \ln(r_\star/\rcore)\right]^{1/2} & r_\star > \rcore = r_{\rm flat}.
\end{array} \right. 
\end{eqnarray}
This  suggests that  if we  fix the central  velocity  dispersion, the
maximum  circular velocity  of the halo  is degenerate with the core radius,
and must increase with core radius
as $\Vmax    \propto \rcore$  in the  large-core   regime. 
Another way to interpret the result is that in this regime, the core
density of the dark matter halo is fixed, but not the core size.
Conversely, in the small core regime we expect  $\Vmax$ to vary slowly
with  increasing  core size  (  $\Vmax     \propto   \sqrt{1  +
2\ln(r_\star/\rcore)}$  for    this example).    Interestingly,  these
qualitative features are demonstrated by our  best-fit contours in the
right panel of Figure \ref{fig:rcore}.

For a more gradual transition from core to flat rotation,
we allow $a < 1$ with $\rcore < r_{\rm flat}$.  In this case, 
for $r_\star < r_{\rm flat}$, we have
\begin{eqnarray}
\sigma_\star^0 = 
 \left\{\begin{array}{cll}
 &  a \Vmax \frac{1}{\sqrt{2}}  \left(\frac{r_\star}{\rcore}\right) &  r_\star \le \rcore \\
 &  a \Vmax  \left[ \frac{r_\star}{\rcore} - \frac{1}{2} \right]^{1/2} & r_\star > \rcore.
   \end{array} \right.
\end{eqnarray}
This means that  for a fixed core size,  a weaker transition  (smaller
$a$)  requires  a larger $\Vmax$  to   reproduce a given $\sigma_*^0$:
$\Vmax \propto \rcore/a$ or fixed halo core density in  the
large-core regime, and $\Vmax \propto 
\sqrt{\rcore}/a$  in the   small-core  regime ($\rcore  \ll r_\star$).
This result demonstrates qualitatively why the implied core
sizes increase with a sharper transition in the assumed dark matter
profile ($\alpha:  1  \rightarrow 2$ in
equation \ref{eqt:rho}).   The stellar velocity dispersion  is roughly
governed   by $V_c(r  \simeq   r_\star)$.   As the  transition becomes
sharper, the ratio $V_c(r = r_\star)/\Vmax$ increases, and an observed
stellar velocity dispersion can be produced with a smaller $\Vmax$.

The  scalings seen  in  the left-hand  panel  of   Figure \ref{fig:rcore}  for  $Q =
\rho_0/\sigma_{\rm dm}^3$ vs. $\Vmax$ can be similarly understood.  In
the large-core regime, where   $r_\star \lesssim \rcore$,  the central
density $\rho_0$ will   be  approximately constant    because   fixing
$\sigma_\star^0$ roughly fixes $V_c(r_\star)$.  
In this case, we expect
$Q   \propto   \sigma_{\rm  dm}^{-3}    \propto    (a\Vmax)^{-3}  \propto
\rcore^{-3}$, which reproduces the contour  scaling in Figure \ref{fig:rcore}.  

The dependence   of $Q$   on the   transition parameter  $a$  is 
weaker  than the  scaling of $\Vmax$: $Q^{-1}
\propto a\Vmax\rcore^2[1 - 2\ln(a)]^{3/2}$.  This result is consistent
with our numerical  findings: the  sharper the transition, 
the larger is the core size required for a given $Q$.
In the small core regime, we have seen that $a\Vmax \propto
\sqrt{\rcore}$ and hence $Q\propto \rcore^{-2.5}$ implying that the
$\rcore$ vs $Q$ contour should steepen slightly for small values of
$\rcore$.   

\section{Results}
\label{sec:results}

\subsection{Phase-Space Constraints}

The derived relationship between allowed $\Vmax$  and  $\rcore$ 
values becomes interesting when expressed in terms
of the implied central   phase-space  density for each
solution. Consider the case  of WDM with a distribution function given by 
\beq
f = g \frac{\beta}{e^{p/T} + 1},  
\label{eq:df} 
\eeq
where $g$ is the number of spin degrees of freedom of the particle. 
For $\beta = 1$ the distribution is thermal, and in this case we define the 
mass of the WDM particle as $m_x$. 
An example of a (non-thermal) $\beta \ne 1$ distribution is WDM from
an oscillation-produced sterile neutrino \citep[e.g.][]{dodelson_widrow94}.
If $\beta$ is independent of momentum, we may calculate the
maximum phase-space density by
fixing the present-day density of dark matter
and calculating the velocity dispersion from equation \ref{eq:df}:
\beq
Q_{\rm max}  = 5 \times 10^{-4} \beta \, \left (\frac{g}{2} \right )\left (\frac{m_x}{1~{\rm keV}} \right )^{4} 
{\rm \Msun\,pc^{-3}(\kms)^{-3}}. 
\label{eq:Q}
\eeq
As discussed in the introduction, the (course-grained) 
phase-space density in any
WDM galaxy halo can never exceed $Q_{\rm max}$.

Measurements of  the Ly$\alpha$ power spectrum   place a limit  on the
free streaming  scale (or equivalently, the present-day velocity)
of  WDM particles.  We can convert this directly to a
lower limit on $Q$. For a fixed free-streaming  scale,
the  mass of a   thermal and non-thermal  particle is related by
$m_x/T_x =    m_s/T_s$, where $m_s$ is the    mass of  the non-thermal
particle and  $T_s$  is its  corresponding temperature  in   equation
\ref{eq:df}.  Fixing the present-day density of dark matter
in this case implies  
$\beta   = (m_s/m_x)^4$ as long as $\beta$ is independent of
momentum.  Note that since $\beta$ and $Q$ both scale as the fourth
power of the WDM particle mass,
any  Ly$\alpha$ forest constraint can be mapped uniquely 
to a $Q_{\rm max}$ constraint, independent of the particle mass.

The most conservative quoted constraint from the Ly$\alpha$ forest
 power spectrum expressed as a limit on the
thermal WDM particle mass is $m_x \gtrsim~0.5$ keV
\citep{viel_etal05,abazajian06a}.  A 
distinct analysis   using different simulations  has   derived an even
stronger limit $m_x
\gtrsim~2.5$  keV \citep{seljak06}.  The 
$m_x \gtrsim~0.5$ keV limit
implies an upper limit  on the phase-space  density in the core of any
galaxy of $Q \gtrsim 3
\times 10^{-5} {\rm \Msun\,pc^{-3}(\kms)^{-3}}$.  That is, if WDM accounts for the mass of 
the dark halo of Fornax, then  Figure \ref{fig:rcore} implies that the  core must be quite small,
$\rcore \lesssim  85$  pc,  and  $\Vmax \lesssim~35~\kms$  (the  lower
dotted    line).    The more    stringent 
$m_x \gtrsim 2.5$ keV   constraint 
implies    $Q       \gtrsim       0.02~{\rm
\Msun\,pc^{-3}(\kms)^{-3}}$ and  demands  $\rcore \ll 10~$pc even  with
allowances for extremely sharp transitions to the profile core.

Of   course,  the Ly$\alpha$  forest constraint   only  applies to the
specific case of a WDM particle with  a phase-space density given by
equation \ref{eq:Q}.  More generally, there  are models with a finite
primordial    phase-space density that do    not have the same mapping
between $Q$  and $P(k)$ as  does WDM.  For example, the well-motivated
SuperWIMP  scenario  suggests that  dark matter
arises  from out-of-equilibrium decays   and   this gives  rise  to  a
non-thermal spectrum
\citep{cembranos_etal05,kaplinghat05}.  Also,
non-thermal resonantly-produced sterile  neutrino dark matter would be
"cooler" than a thermal or non-resonant sterile WDM candidate
\citep{shi_fuller99,abazajian_etal01}.
 Yet another example is fuzzy dark matter \citep{hu_etal00}.

In cases such as these the relation between the phase-space density and the power spectrum is different from that implied by 
equation \ref{eq:Q}, a more  direct  observational constraint on  $Q$ may  be  derived from
considering the rotation curves of low-mass spiral galaxies
\citep{dalcanton_hogan01,kormendy_freeman04}. 
Among the highest-resolution rotation curves  for low-mass spirals are
those  presented  by  \citet{simon_etal05}, who  used  two-dimensional
H$\alpha$  velocity fields    to  extract rotation   curves for   five
galaxies.  When the dark matter halo components  of these galaxies are
fit to cored density distributions  the central densities range from $
\rho_0 \simeq  0.1 - 0.5\ {\rm  M_{\odot}\,pc^{-3}}$  and the rotation
curves   flatten at   $\Vmax    \gtrsim  74-114~\kms$.   If we   adopt
$\sigma_{\rm dm} \simeq 0.55  \Vmax$ (as is  appropriate for the cored
profiles we   have  used here)  then  the implied  central phase-space
limits are $Q \gtrsim$ ($1.7$, $4.4$, $1.7$, $1.6$, and $0.5$) $\times
 10^{-6}{\  \rm    \Msun\,pc^{-3}(\kms)^{-3}}$ for  NGC2976,  NGC4605,
NGC5949, NGC5963,  and NGC6689, respectively.   If we (conservatively)
adopt  $Q  \gtrsim  10^{-6}{\   \rm \Msun\,pc^{-3}(\kms)^{-3}}$ as   a
(direct) constraint on  $Q$, this provides  a fairly clean upper-limit
on the Fornax core radius, $\rcore \lesssim 300$~pc, for a broad class
of  dark  matter   candidates      (upper  dotted line      in  Figure
\ref{fig:rcore}).

\subsection{Phase-Space Independent Constraints}

While the most popular alternatives to CDM
produce cores because of their phase-space constraints,
 cores in dark  matter halos need not  arise as a result of
large   velocity    dispersions.
For example, self-interaction or self-annihilation would produce  cores even if
the             dark          matter               is             cold
\citep{spergel_steinhardt00,medvedev00,kaplinghat00,sanchezsalcedo03}.
Astrophysical or dynamical effects could also produce cored profiles.
For our most general constraint we can
use the $\rcore$-$\Vmax$ relation demanded by the velocity
dispersion data to place a reasonable  upper limit on the core size in 
Fornax by imposing a plausible upper limit on $\Vmax$ for 
its halo.    

Models of Fornax with large $\Vmax$  have large masses. As a result, they will experience 
 large dynamical friction and spiral to the galactic center very fast. Therefore only in the rare circumstance of a 
very recent accretion event could a massive subhalo at a distance
of $\sim 140$ kpc from the Milky Way be possible.  \cite{zentner_03}
found that only $5\%$ of Milky Way-sized~\LCDM~halos contain a subhalo with
$\Vmax > 100~\kms$ and that the fraction falls sharply beyond this
point.  Since the presence of dark matter cores would only decrease
the subhalo population, we may adopt $\Vmax \lesssim 100~\kms$ as
a reasonable upper limit on the Fornax halo.  
This gives
$\rcore \lesssim 700$~pc as  a conservative general upper limit on the
core size of Fornax.

In light of this argument,
a core large enough to explain the extended distribution of globular clusters
in Fornax, $\rcore \simeq 1.5$~kpc, seems highly implausible.
Such a case would demand a very large Fornax
dark halo: $\Vmax \gtrsim 200~\kms$ with $M(< r_{\rm max}) \simeq 4.4 \times 10^{11} \Msun$.  The dynamical friction timescale
for such an object at a distance of $\sim 140$~kpc is less than
$\sim 0.5$Gyr. Consider also the fact that the stellar mass
of Fornax is $M_\star \simeq 3 \times 10^7 \Msun$ \citep{mateo_98}.
This means that a $\sim 1.5$~kpc core in Fornax
 would require {\em both} that the Milky Way
is experiencing a (very rare) 
recent equal-mass merger  {\em and} that
the merging object is a system with a stellar baryon fraction that is
$\sim 10,000$ times smaller than the universal baryon fraction.

\section{Conclusions and Discussion}

In this paper, we have shown that a  standard WDM particle cannot
produce a core  in Fornax larger than $\sim   85$~pc without
saturating  the implied Ly$\alpha$  forest bound on the phase-space
density.  For a more general class of dark matter, the relation between the phase-space density 
 and the power spectrum may be different, and in these
cases the limit on $Q$ comes from  low-mass  galaxy  rotation curves, giving $\rcore \lesssim
300$~pc.   Irrespective of the origin of the core,
dynamical arguments alone place an upper limit on the Fornax halo
mass and demand $\rcore \lesssim 700$~pc.

One important caveat to the above results is that there is no
concrete prediction for the precise shape or normalization of dark
matter halo profiles in cases  where the dark matter   is not cold.
Even  in well-studied  cases like  WDM,   a phase-space  core  has
never  been realized  in an  N-body  simulation  and the   nature of
the  expected transition between a rising $\rho(r)$ profile  and the
flat core region is unknown.  The   density profile we have adopted
in equation \ref{eqt:rho}   asymptotes  to the  CDM    prediction at
large radius, $\rho(r) \propto  r^{-3}$ and allows some freedom  in
the nature of the transition  from the  inner   core region to  the
outer fall-off  in density.   The sharpness of this  transition  is
captured in the shape parameter $\alpha$: the sharper  the transition
(larger $\alpha$), the closer the core radius is to the  radius of maximum rotation.
Given our  general  ignorance as to   the ``correct'' shape,   we have
explored the degree   to which our   constraints on the core  size
change when we  adopt $\alpha=1$ and  $2$.   The $Q$ constraint for  a
broad  class of dark matter models  derived  from from low-mass spiral
galaxies yields $\rcore \lesssim 500$~pc and $200$~pc for $\alpha = 2$
and  $1$, respectively.  For WDM, the Ly$\alpha$ forest power
spectrum implies $\rcore \lesssim 150$~pc  and $40$~pc, for $\alpha  =
2$ and $1$, respectively.

What do our results mean for the globular  clusters in Fornax?  A core
large  enough   to  explain  the   extended  distribution of  globular
clusters,  $\rcore  \gtrsim 1.5$~kpc,   would   imply  a   dynamically
implausible dark  halo, $\Vmax \gtrsim 200~\kms$ (see Figure \ref{fig:rcore}).  
In addition, such a core is vastly inconsistent with any viable WDM model. 

Given these constraints, an alternative explanation for the globular cluster system may be considered. 
For example, the profile could be cuspy, but this would require the globular clusters to have formed 
at large radii beyond the asymptotic central slope (J. Read, private communication). 
This seems to require a fine-tuned timing argument. The timing argument could be sightly alleviated 
by arguing that only the innermost globular cluster defines the core ($\rcore \gtrsim 240$ pc), however
even this case is inconsistent with standard WDM. 

We may also consider a solution based on tidal heating. For  example, dynamical
friction effects on  the orbital decay of  the globular clusters can  be counteracted by
dynamical   heating  via tidal shocks    or by relaxation during major
mergers  \citep{oh_etal00}.  The tidal   heating explanation  may have
difficulty because the   proper motion measurements indicate that  the
Fornax is currently  near its orbital  pericenter and has apocenter of
$\approx 250$~kpc
\citep{dinescu_etal04} and thus does not come sufficiently close to
the center of the Milky Way to experience strong tidal force.  Note,
however, that the tidal heating could have been provided by another
object in the past and not necessarily by the Milky Way progenitor \citep[see,
e.g.,][]{kravtsov_etal04}. Heating by a major merger with another
dwarf galaxy is plausible, as there is evidence for such recent
major merger in Fornax from observations of extra-tidal stars
\citep{coleman_etal05,olszewski06}. Although such mergers are not
likely for satellites orbiting in a potential of a much larger host
halo, given large pericentric and apocentric distances of the Fornax,
it is likely that it has been accreted by the Milky Way only
recently. Major mergers are much more likely prior to accretion by a
large host. Other possible, albeit more speculative, explanations,
such as core formation due to inspiralling of multiple black holes,
are discussed in \citet{sanchezsalcedo_etal06}.

As     emphasized       by         \citet{goerdt_etal06}           and
\citet{sanchezsalcedo_etal06},  the    Fornax globular  cluster system
presents  an    interesting  case study    with  possible  fundamental
implications for  the   nature of  dark  matter.  Here   we argue that
interpretation in terms of  large dark matter  core is problematic, as
such a  large   core is  not   consistent  with   popular dark   matter
alternatives to  CDM. Large  cores also appear to be inconsistent
with high-resolution  measurements  of  rotation  curves of   the  LSB
galaxies for an even broader class of dark matter scenarios.  
Dynamical heating of  the   globular clusters by tides   or
relaxation in mergers  could  provide an  alternative  explanation and
needs to be explored. 

\acknowledgements

We are grateful to Terry Walker and the Center for Cosmology and
Astro-Particle Physics (CCAPP) at The Ohio State University for
hosting the workshop on alternative dark matter models in January of
2006, where this work was initiated.  We thank Ben Moore, Justin Read, and 
Andrew Zentner for enlightening discussions. LES is supported in part by a
Gary McCue Postdoctroral Fellowship through the Center for
Cosmology at UC Irvine.  
AVK is supported by the NSF grants AST-0206216, AST-0239759, and
AST-0507596, by NASA through grant NAG5-13274, and by the Kavli
Institute for Cosmological Physics at the University of Chicago.  
KA is supported by Los Alamos National Laboratory under DOE contract W-7405-ENG-36.
AK is supported by NSF grant AST-0407072. We acknowledge Larry's for inspiration.

\bibliography{fornax}

\end{document}